\documentclass[aps,nofootinbib,preprint]{revtex4-1}

\usepackage{verbatim}
\usepackage[T1]{fontenc}
\usepackage[utf8]{inputenc}
\usepackage[american]{babel}
\usepackage{epsfig}
\usepackage{graphicx,subcaption,caption}
\usepackage{booktabs}
\usepackage{multirow}
\usepackage{dcolumn}
\usepackage{amsmath}
\usepackage{mathtools}
\usepackage{amsfonts}
\usepackage{amssymb}
\usepackage{ulem}
\usepackage{epstopdf}
\usepackage{bm}
\usepackage{siunitx}
\usepackage{braket}
\usepackage{enumitem}
\usepackage{soul}
\usepackage[table]{xcolor}
\usepackage{color}
\usepackage{transparent}
\usepackage{pifont}
\usepackage{enumitem}
\usepackage{CJKutf8}

\definecolor{navyblue}{rgb}{0.0, 0.0, 0.5}
\definecolor{royalblue}{rgb}{0.25, 0.41, 0.88}
\definecolor{cadmiumgreen}{rgb}{0.0, 0.42, 0.24}
\definecolor{blue-violet}{rgb}{0.54, 0.17, 0.89}
\definecolor{darkviolet}{rgb}{0.58, 0.0, 0.83}
\definecolor{orange(colorwheel)}{rgb}{1.0, 0.5, 0.0}

\usepackage{hyperref}
\hypersetup{
    colorlinks=true, 
    linkcolor=royalblue, 
    citecolor=magenta}

\usepackage{booktabs}
\usepackage{multirow}
\usepackage{dcolumn}
\usepackage{colortbl}

\begin{document}

\title{
Constraints on the varying electron mass and early dark energy
in light of ACT DR6 and DESI DR2 and the implications for inflation
}
\author{Yo Toda}
\email{y-toda@particle.sci.hokudai.ac.jp}
\affiliation{Department of Data \& Innovation, Kochi University of Technology,  
Tosayamada 782-8502, Japan \looseness=-1}

\author{Osamu Seto}
\email{seto@particle.sci.hokudai.ac.jp}
\affiliation{Department of Physics, Hokkaido University, 
Sapporo 060-0810, Japan \looseness=-1}

\begin{abstract}
Primarily motivated by the Hubble tension, we analyze the varying electron mass model and axionlike early dark energy model (EDE) using baryon acoustic oscillation data from DESI DR2 data and including the recent results from ACT DR6. Our analysis indicates that $m_{e} / m_{e0} = 1.0078 \pm 0.0047$ in the varying $m_e$ model, $m_{e} / m_{e0} = 1.0034 \pm 0.0050$ and $\alpha / \alpha_{0} = 1.0039 \pm 0.0016$ in the varying $m_e$+$\alpha$ model, and the energy fraction of EDE is constrained as $f_\mathrm{EDE} < 0.014$. Since those cosmological models fit with different spectral index $n_s$, we show the posterior of those models on the ($n_s-r$) plane and point out that, for example, Starobinsky inflation works for varying electron mass model while the standard supersymmetric hybrid inflation is preferred in the EDE model.
\end{abstract}
\preprint{EPHOU-25-015}

\maketitle

\section{Introduction}
\label{sec:introduction}

The concordance cosmological model, also known as the $\Lambda$CDM model with cosmological constant $\Lambda$,  is successful for explaining the properties of our Universe with only six parameters; the density parameters of baryon $\omega_b =\Omega_b h^2$ and of cold dark matter (CDM) $\omega_c =\Omega_c h^2$, the angular size of the sound horizon $\theta_\mathrm{MC}$, the optical depth $\tau$, the amplitude of the density perturbation $A_s$ and the spectral index of the scalar perturbation $n_s$, where $h$ is dimensionless Hubble parameter defined as $H_0 = 100 h~\mathrm{km\,s^{-1}Mpc^{-1}}$ with $H_0$ being the present Hubble parameter. Those cosmological parameters have been well evaluated from the measurement of temperature anisotropy and lensing effects in the cosmic microwave background radiation (CMB) by WMAP~\cite{WMAP:2003elm} and Planck~\cite{Planck:2013pxb,Planck:2015fie,Planck:2018vyg}.
In addition to the Planck, Atacama Cosmology Telescope (ACT) is able to measure the CMB anisotropy at much higher multipole $l$ than the Planck and recently its new result data release 6 (DR6) was announced~\cite{ACT:2025fju,ACT:2025tim}.
To resolve the degeneracy between cosmological parameters in the CMB measurements only, the measurement of the baryon acoustic oscillation (BAO) is very powerful. 6dF~\cite{Beutler:2011hx} as well as SDSS~\cite{Ross:2014qpa,BOSS:2016wmc} had measured BAO, and combined analysis of CMB and BAO have been performed with data from those and the Planck, and show consistency of the $\Lambda$CDM model~\cite{Planck:2018vyg,ACT:2025fju,ACT:2025tim}. 

Recently, some estimation of cosmological parameters appear to have tensions between them.
First, $H_0$ is estimated as $H_0 \simeq 68~\mathrm{km\,s^{-1}Mpc^{-1}}$ from CMB and BAO measurement,
 while the local measurement by, for instance, SH0SE~\cite{Breuval:2024lsv} and HOLiCOW~\cite{H0LiCOW:2019pvv} have reported a larger value. This discrepancy is referred to as the Hubble tension that a lot of attention have been paid; see also the latest review~\cite{CosmoVerse:2025txj}.
Second, the results of recent BAO measurement by Dark Energy Spectroscopic Instrument (DESI) survey has released  from DR1~\cite{DESI:2024uvr,DESI:2024lzq,DESI:2024mwx} and DR2~\cite{DESI:2025zpo,DESI:2025zgx}.
These results indicate that the standard cosmological constant dark energy model, characterized by $(w, w_a) = (-1, 0)$, shows a tension at the level of $2.8\sigma - 4.2\sigma$ when combined with DESI BAO, CMB, and SNe data~\cite{DESI:2025zgx}, where $w_a$ denotes the derivative of the equation-of-state parameter $w$ with respect to the scale factor $a$.
Third, the latest results from ACT combined with CMB lensing and BAO indicates a larger spectral index of $n_s = 0.974 \pm 0.003$~\cite{ACT:2025tim, ACT:2025fju} than that from the Planck $n_s \simeq 0.96$. When combined with constraints on the tensor-to-scalar ratio $r$ from the BICEP and Keck data~\cite{BICEP:2021xfz}, the result shows that the Starobinsky inflation model locates at the edge of the $2\sigma$ level.

Primarily motivated by the Hubble tension, in this paper,
 we will consider an extended model of cosmology.
The varying electron mass model is regarded as one of the most promising approaches to alleviate the Hubble tension~\cite{Schoneberg:2021qvd}.
A larger electron mass during recombination leads to earlier recombination, resulting in a shorter sound horizon and thus inferred a larger $H_0$.
This model has been extensively studied from multiple perspectives~\cite{Hoshiya:2022ady, Seto:2022xgx, Seto:2024cgo, Toda:2024ncp, Toda:2024uff, Toda:2025dzd, Sekiguchi:2020teg, Sekiguchi:2020igz, Hart:2017ndk, Hart:2019dxi, Chluba:2023xqj, Smith:2018rnu, Lynch:2024hzh, Schoneberg:2024ynd, Khalife:2023qbu, Zhang:2022ujw, Greene:2023cro, Greene:2024qis} and motivated by the theory that has the couplings of the matter fields to additional scalar fields~\cite{Carroll:1998zi,Brax:2002nt,Chiba:2006xx, Baryakhtar:2024rky}.
Another interesting model where shorter sound horizon can be realized is 
Early dark energy (EDE)~\cite{Poulin:2018dzj, Poulin:2018cxd, Braglia:2020bym, Agrawal:2019lmo, Ye:2020btb, Smith:2019ihp, Lin:2019qug, Niedermann:2019olb, Niedermann:2020dwg, Seto:2021xua}, in which the dark energy density temporarily becomes non-negligible only at recombination epoch, also reduces the sound horizon. As a result, a larger $H_0$ is inferred.
In our previous studies~\cite{Seto:2024cgo,Toda:2025dzd}, we have shown that an increased electron mass or the presence of EDE seems to be consistent with the DESI BAO measurement, and in particular the analysis including DR2 data indicates that a larger electron mass is favored at more than the $2\sigma$ level~\cite{Toda:2025dzd}.
ACT Collaboration also have examined EDE models and the varying electron mass model
 combined with DESI DR1 data and found the energy fraction of EDE is constrained as $f_\mathrm{EDE} < 0.012$ and no variation of electron mass is consistent within uncertainty~\cite{ACT:2025fju}.

In this paper, we examine those scenarios including the latest CMB data from the Atacama Cosmology Telescope (DR6)~\cite{ACT:2025tim, ACT:2025fju} with DESI DR2 data.
In addition, we discuss the implication on inflation model for those cosmological model, since
 EDE models tend to prefer a larger value of $n_s$~\cite{Jiang:2022uyg,Ye:2022efx,Peng:2023bik}
 and the varying electron mass model indicates a smaller spectral index $n_s$~\cite{Sekiguchi:2020teg}.

This paper is organized as follows. We present the explanation of the model in Sec~\ref{sec:models}, the method and the datasets of our analysis in Sec.~\ref{sec:data}, the results in Sec.~\ref{sec:results}, and the summary in Sec~\ref{sec:conclusions}.

\section{Models}
\label{sec:models}

\subsection{Varying electron mass}

We review the effects of varying electron mass during recombination with comparing it to our previous research~\cite{Toda:2025dzd}. 

In this scenario, it is assumed that the electron mass at the era of recombination
 differs from its current value ($m_{e0} = 511\,\mathrm{keV}$) and became the present value
 after the recombination had completed so that it is consistent with constraints from quasar spectra~\cite{Uzan:2024ded, Murphy:2016yqp, Evans:2014yva, Murphy:2017xaz, Songaila:2014fza, AlbornozVasquez:2013wfw, Dapra:2015yva, Bagdonaite:2015kga}.
 In our modification of \texttt{camb}, this transition is implemented as a step function with a threshold at redshift $z = 50$\footnote{We choose the transition redshift $z = 50$ because it is sufficiently late for recombination to have been completed, and early enough to precede the formation of the first stars.}, and we did not assume a specific dynamics in this paper\footnote{When we consider the varying fine structure constant $\alpha$ model in our paper, it is also implemented as a similar step function.}.

A larger electron mass increases hydrogen energy levels and the energies $E$ (or frequencies $\nu$) of Lyman-alpha photons, because $E$ and $\nu$ scale as $E = (m_{e}/m_{e0}) E_0$ and $\nu = (m_{e}/m_{e0}) \nu_0$.
Here, $m_e (m_{e0})$ is the electron mass at the recombination (present) epoch,
and $E_0$ and $\nu_0$ are the values for $m_{e0}$.
Thus, a larger energy is required to excite hydrogen for a larger $m_e$.
As a photon energy decreases by the redshift ($E=\nu \propto 1/a$ with $a$ being the scale factor), 
 the photon energy becomes smaller than the energy for the ionization of hydrogen
 earlier than the time in the case of constant electron mass of $m_{e0}$, 
 which causes the earlier recombination of hydrogen.
Then, the sound horizon at the recombination is shortened as
\begin{equation}
r_s = \int_{0}^{t_*} \frac{c_s(t)}{a(t)} dt,
\end{equation}
where $t_{*}$ is the last scattering time and $c_s$ is the sound speed. 
The diameter distance is also shorter to keep the observed angular scale of CMB acoustic peaks,
which infers a larger $H_0$.

A larger electron mass also affects the Silk damping scale $\lambda_D$ whose squared is given by
\begin{equation}
\lambda_D^2 = \frac{1}{6} \int_{0}^{\eta_\mathrm{dec}} \frac{d\eta}{\sigma_T n_e a} 
\Bigg[ \frac{R^2 + \frac{16}{15}(1+R)}{(1+R)^2} \Bigg],
\end{equation}
because the Thomson cross section $\sigma_T = \sigma_{T,0} (m_{e0}^2 / m_e^2)$
 decreases as the electron mass increases.
Here, $\sigma_{T,0}$ is the Thomson scattering cross section for the case that the value of the electron mass is same as $m_{e0}$, $n_e$ is the free electron density, $R = 3\rho_b / 4\rho_r$ is the baryon-to-radiation ratio, and $\eta_\mathrm{dec}$ is the decoupling conformal time. 
The ratio of the Silk damping scale to the sound horizon is unchanged under the variation of
 the electron mass nevertheless~\cite{Sekiguchi:2020teg}.
We also take account of other minor effects such as changes in photoionization cross sections, recombination and ionization rates, $K$ factors, Einstein $A$ coefficients, and two-photon decay rates. See Ref.~\cite{Planck:2014ylh} for details of those effects.

Alleviation of the Hubble tension in the varying electron mass model is achieved by simultaneous increased values of $\omega_\mathrm{c}$ and $\omega_\mathrm{b}$ along a parameter degeneracy. 
This, in turn, leads to a larger $\theta_d(z) \equiv r_d / D_M$, where the sound horizon is defined as $r_d = \int_{z_d}^{\infty} \frac{c_s(z)}{H(z)} dz$, the comoving angular diameter distance is $D_M(z) = \frac{c}{H_0} \int_0^z \frac{dz'}{H(z')/H_0}$, and $z_d$ is the redshift of the drag epoch.
Thus, this parameter degeneracy can be broken by taking BAO data into account~\cite{Sekiguchi:2020teg}.

We also note that the joint analysis of the latest CMB, BAO, and SNe Ia observations constrains
 the deviation of the electron mass.
The likelihood on $m_e$ somewhat depends on which BAO data are included in analysis.
In the previous work~\cite{Toda:2025dzd} where we have used Planck data for CMB,
 we have found  
\begin{equation}
m_e/m_{e0} = 1.0101 \pm 0.0046,
\end{equation}
 for the DESI BAO DR2, and 
\begin{equation}
m_e/m_{e0} = 1.0049 \pm 0.0055,
\end{equation}
 for the 6dF and SDSS BAO data~\cite{Beutler:2012px, Ross:2014qpa, Alam:2020sor}, respectively.
$m_e/m_{e0}=1$ is more than $2\sigma$ away in the former, while it is consistent in the latter.  
We will see how those change if we include ACT data for CMB in our analysis.

\subsection{Early Dark Energy}

The potential of the axionlike EDE takes the form~\cite{Poulin:2018dzj}
\begin{equation}
V (\phi) = \Lambda^4 \left( 1 - \cos\left(\frac{\phi}{f}\right) \right)^n \,,
\end{equation}
where $\Lambda$ is the energy scale of the potential, $f$ is the breaking scale of the shift symmetry, and $n$ is the power index of the cosine function and we set $n=2$ in this paper.
As in the previous paper~\cite{Poulin:2018cxd}, we use the three phenomenological parameters: $z_c$, $\Theta_i$, and $f_\mathrm{de}(z_c)\equiv\frac{\rho_\mathrm{de}(z_c)}{\rho_\mathrm{tot}(z_c)}$, which stand for the redshift when $\phi$ starts to oscillate, the initial value of the scalar field $\phi/f$, and the energy fraction of EDE to the total energy density $\rho_\mathrm{tot}$ at $z_c$. 

In this model, the EDE component becomes significant near the transition redshift $z_c$, contributing several percent to the total energy density. 
This results in a larger Hubble parameter around $z_c$ and a shorter sound horizon, which leads to a higher present Hubble constant.
After the transition ($z < z_c$), the energy density of EDE decreases as $\rho_\mathrm{de} \propto a^{-4}$ for $n = 2$, which is faster than the scaling of the background energy density.

To examine the axionlike EDE, we use the \texttt{camb}~\cite{Howlett:2012mh} where axionlike EDE is already implemented and perform the MCMC analysis, sampling $f_{\rm de} (z_c)\in[0.00001,0.15]$, $z_c\in[1000,50000]$, and $\Theta_i\equiv \phi_{\rm ini} / f \in[0.01,3.14]$ in addition to the 6 standard parameters.

\section{datasets and methodology}
\label{sec:data}

We perform an MCMC analysis of the varying electron mass model and the EDE model using the public MCMC code \texttt{Cobaya}\footnote{\texttt{Cobaya} and the cosmological data are available at \url{https://github.com/CobayaSampler.}}~\cite{Torrado:2020dgo},
requiring the convergence criterion $R - 1 < 0.02$ ($R - 1 < 0.03$ only for P-ACT-BK-LB2S on EDE).
We also use the cosmological Boltzmann code \texttt{camb}~\cite{Lewis:1999bs,Howlett:2012mh}
and the recombination code \texttt{recfast}~\cite{Scott:2009sz}, with the relevant modifications implemented.

We use the following datasets:
\begin{itemize}

\item \textbf{ACT CMB:} The temperature and polarization likelihoods for high $l$ from ACT DR6~\cite{ACT:2025fju}. We also include the Planck Sroll2~\cite{Pagano:2019tci} likelihood for low-$l$ polarization and CMB lensing from ACT~\cite{ACT:2023kun,ACT:2023dou}.

\item \textbf{ACT CMB with Planck cut:} In addition to the ACT CMB data, we include a ``$\mathrm{Planck_{cut}}$'' dataset using Planck high-$l$ data for $l < 1000$ in TT and $l < 600$ in TE/EE~\cite{Planck:2019nip}.

\item \textbf{BICEP and Keck:} Measurements of CMB B-modes~\cite{BICEP:2021xfz}.

\item \textbf{Type Ia Supernovae:} Light curves from \texttt{Pantheon+}~\cite{Pan-STARRS1:2017jku}.

\item \textbf{BAO:} Distance measurements from DESI DR2~\cite{DESI:2025zgx}.

\item \textbf{$M_B$:} Local $H_0$ measurement by Riess et al~\cite{Riess:2020fzl}, in terms of magnitude measurement
\footnote{
This prior shifts several cosmological parameters in a way that increases the Hubble constant. Therefore, it is introduced solely for the purpose of evaluating how well a given model fits the data, including the SH0ES measurement, in comparison with the $\Lambda$CDM model.
}
.
\end{itemize}

We refer to these data combinations as shown in Tab.~\ref{Tab:data}.
\begin{table}[h]
\centering
\begin{tabular}{|c|l|}
\hline 
ACT-LB2S & ACT CMB, BAO, and SNIa \\ \hline
P-ACT-LB2S & ACT CMB with Planck cut, BAO, and SNIa \\ \hline
P-ACT-LB2SMb & ACT CMB with Planck cut, BAO, SNIa, and $M_B$ \\ \hline
P-ACT-BK-LB2S & ACT CMB with Planck cut, BICEP-Keck, BAO, and SNIa \\ \hline
\end{tabular}
\caption{Summary of dataset combinations used in this analysis.\label{Tab:data}}
\end{table}

\section{Results and Discussion}
\label{sec:results}
We present the results of our analysis. The one-dimensional and two-dimensional marginalized posteriors for various cosmological parameters are shown in Figs.~\ref{fig:me} and~\ref{fig:EDE}, and the corresponding 68\% confidence level constraints are summarized in Table~\ref{table}.

In Table~\ref{table}, we list the 68\% confidence level constraints for selected cosmological parameters, along with their Gaussian tensions.
The Gaussian tension for $H_{0}$ is calculated as
\begin{equation}
T_{H_{0}} = 
\frac{H_{0~\mathrm{\mathcal{D}+DESI-DR2}} - 73.17}{
\sqrt{\sigma_{\mathrm{\mathcal{D}+DESI-DR2}}^{2} + 0.86^{2}}}\,,
\label{eq:TH0}
\end{equation}
quantifying the Hubble tension relative to the SH0ES measurement~\cite{Breuval:2024lsv}, and the tension for $\sigma_{8}$ is given by
\begin{equation}
T_{\sigma_{8}} = 
\frac{\sigma_{8~\mathrm{\mathcal{D}+DESI-DR2}} - 0.802}{
\sqrt{\sigma_{\mathrm{\mathcal{D}+DESI-DR2}}^{2} + \frac{0.022^{2} + 0.018^{2}}{2}}}\,,
\end{equation}
quantifying the tension with the Kilo-Degree Survey (KiDS-Legacy) result~\cite{Stolzner:2025htz}.
Here, $H_{0~\mathrm{\mathcal{D}+DESI-DR2}} $ in Eq.~(\ref{eq:TH0}) is in the unit of km$/$s$/$Mpc.
It is worth noting that the latest KiDS-Legacy estimate, $\sigma_{8} = 0.802^{+0.022}_{-0.018}$, lies between various other measurements of $\sigma_8$ and is relatively optimistic.

In the Tab.~\ref{table-best}, we also summarize the best-fit values of the cosmological parameters and their corresponding minimized chi-squared values. Increasing the number of model parameters typically leads to an improvement in the chi-squared value and direct comparison of $\chi^2_{\mathrm{min}}$ values is not meaningful. Then, to penalize increasing the number of parameters and facilitate fair comparison, we also calculate the Akaike Information Criterium (AIC) of model $X$ relative to that of $\Lambda$CDM model as follows:
\begin{equation}
    \Delta AIC = \chi^2_{\mathrm{min}, X} -\chi^2_{\mathrm{min},\Lambda\mathrm{CDM}}
    + 2(N_{X}-N_{\Lambda \mathrm{CDM}}),
\end{equation}
where $N_{X}-N_{\Lambda \mathrm{CDM}}$ is the number difference in the free parameters between the model $X$ and $\Lambda$CDM model.

\subsection{Varying electron mass}

We find that a slightly larger electron mass, $m_e / m_{e0} = 1.0174 \pm 0.0065$ (ACT+LB2S), is preferred, with the standard value $m_e / m_{e0} = 1$ lying more than $2.5\sigma$ away.
The height of the first acoustic peak of the CMB power spectrum is not normalized by ACT CMB data alone.
In fact, the ACT+LB2S outcome is dominantly constrained by the DESI BAO data, leading to higher inferred values of both $H_0$ and $m_e$.
When the Planck cut is applied, the preferred electron mass shifts closer to the standard value, yielding $m_e / m_{e0} = 1.0078 \pm 0.0047$.
We also quote $m_e / m_{e0} = 1.0101 \pm 0.0046$ (Planck+LB2S) from Ref.~\cite{Toda:2025dzd}, and it is worth noting that this $\sim 1\%$ preference for a larger electron mass remains supported when considering ACT data, which provide high-precision measurements of the CMB power spectrum in the high-$l$ multipole range.
Here, the reason P-ACT+LB2S indicating the smallest value of $m_e/m_{e0}$ can be understood as follows.
The P-ACT+LB2S analysis utilizes CMB data over a wider multipole range, $0 < l < 4000$, compared to ACT-LB2S and Planck-LB2S.
This broader coverage limits the extent to which parameter degeneracies can operate freely, resulting in a slightly smaller value of $m_e / m_{e0}$ in the P-ACT+LB2S analysis.

As discussed in the ACT analysis~\cite{ACT:2025tim}, the P-ACT-only dataset does not favor a larger electron mass on its own. 
In contrast, the preference for a larger electron mass primarily comes from the DESI BAO data.
The DESI BAO results show that the sound horizon relative to the angular diameter distance, $r_d / D$, is slightly larger than predicted by the standard $\Lambda$CDM model, as shown in Fig.~6 of the DESI paper~\cite{DESI:2025zgx}.
A measured increase in $r_d / D$ implies a larger product of the sound horizon and the Hubble constant, $r_d H_0$, since the angular diameter distance $D$ is inversely proportional to $H_0$.
In the varying electron mass model, a larger electron mass slightly increases the product $r_d H_0$ and becomes consistent with the DESI BAO results (see also Ref.~\cite{Seto:2024cgo} for a detailed discussion).

\begin{figure}[ht]
\includegraphics[width=17cm]{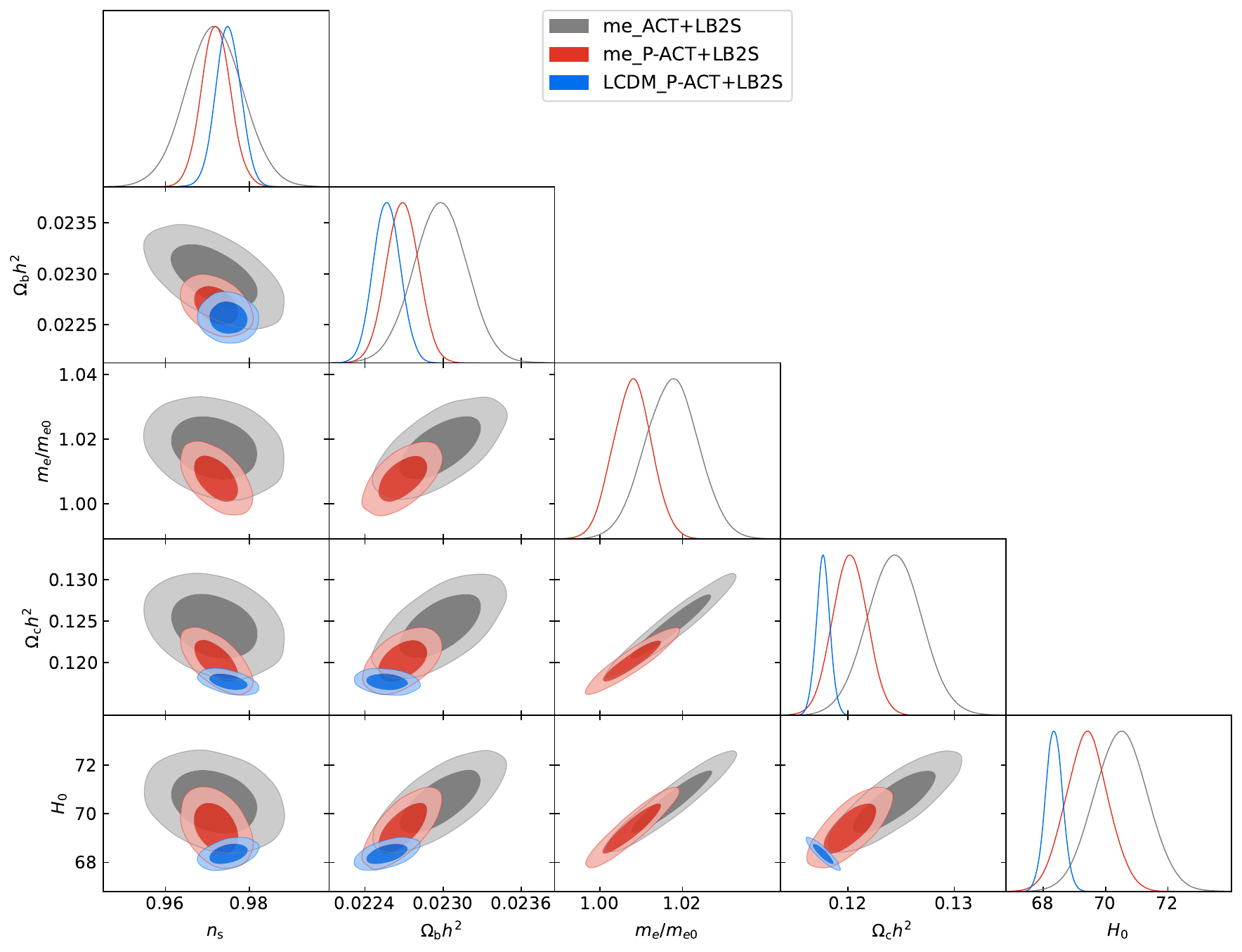} 
\centering
\caption{Posterior distributions of selected parameters for the varying $m_e$ model. Gray and red contours correspond to the different datasets indicated in the legend, while blue contours represent the $\Lambda$CDM baseline for comparison.}
\label{fig:me} 
\end{figure}

As discussed in Sec.~\ref{sec:models}, the varying electron mass model can relax the Hubble tension without significantly compromising the CMB fit.
This can also be understood as parameter degeneracies among the cold dark matter density $\omega_\mathrm{c} = \Omega_\mathrm{c}h^2$, baryon density $\omega_\mathrm{b} = \Omega_\mathrm{b} h^2$, Hubble constant $H_0$, spectral index $n_s$, and the electron mass $m_{e} / m_{e0}$.
From Fig.~\ref{fig:me}, we find the approximate relation for their fractional changes:
\begin{equation}
    \frac{\Delta m_e}{m_{e0}} \simeq
    \frac{\Delta \omega_\mathrm{b}}{\omega_\mathrm{b}} \simeq
    \frac{1}{2} \frac{\Delta \omega_\mathrm{c}}{\omega_\mathrm{c}} \simeq
    \frac{1}{2} \frac{\Delta H_0}{H_0} \simeq
    -2 \frac{\Delta n_s}{n_s}.
    \label{eq.mefit}
\end{equation}

Although the direct impact of the varying electron mass on the damping scale is small, adjusting $\omega_\mathrm{b}$ is necessary to compensate for this effect, which explains the first equality.
To simultaneously match the angular scales measured by CMB and BAO, variations in $\omega_\mathrm{c}$ and $H_0$ are also required, leading to the second and third equality.
These coefficients differ from the theoretical expectations based solely on CMB discussed in Ref.~\cite{Sekiguchi:2020teg} (see also Refs.~\cite{Toda:2024ncp,Jedamzik:2020zmd,Vagnozzi:2023nrq,Pedrotti:2024kpn} for the role of $\omega_c$ to solve the Hubble tension without spoiling the BAO fitting).
In particular, too large $\omega_c$ leads to suppression of the first peak of the power spectrum~\cite{Hu:2001bc}. 
Consequently, a small shift in the spectral index $n_s$ is also needed to maintain a good CMB fit, which explains the final equality.
This is illustrated in Fig.~\ref{fig:TT},\ref{fig:TE}, which shows the difference in the CMB power spectrum relative to the $\Lambda$CDM model.
The upper green curve shows the case with only a $2\%$ increase in electron mass. 
The blue curve additionally includes variations in $H_0$, $\omega_c$, and $\omega_b$ and the red curve further includes a variation in $n_s$ according to Eq.~(\ref{eq.mefit}).

In Tab.~\ref{table}, from the point of view of the Hubble tension, the varying electron mass model is an effective solution, reducing the tension to $3.43\sigma$, which is $1.92\sigma$ lower than that of the $\Lambda$CDM model. 
The best-fit total chi-squared value, $\chi^2_{\mathrm{total}}$, is improved by 7.32, and $\Delta AIC=-5.32$, indicating that this model effectively alleviates the Hubble tension.
Regarding the $\sigma_8$ tension, although the value is slightly worse in the varying electron mass model, the tension is only $1\sigma$.
The obtained constraint on $\Omega_{\mathrm{m}}$ is also consistent with the KiDS result, which reported $\Omega_{\mathrm{m}} = 0.307 \pm 0.011$.

As mentioned above, the preference for a larger electron mass in this results is primarily due to the DESI BAO data.
For comparison, if we use 6df and SDSS instead of DESI BAO, we obtain
\begin{align}
\begin{cases}
m_{e} / m_{e0} = 1.0022 \pm 0.0054 \\
H_{0} = 68.09 \pm 0.95~\mathrm{km\,s^{-1}Mpc^{-1}} \\
n_s = 0.9714 \pm 0.0034
\end{cases} \quad (\mathrm{P\!-\!ACT\!-\!LB_{6S}S}). 
\end{align}
which indicates $m_e/m_{e0}=1$ is well consistent.

\begin{figure}[ht]
\includegraphics[width=17cm]{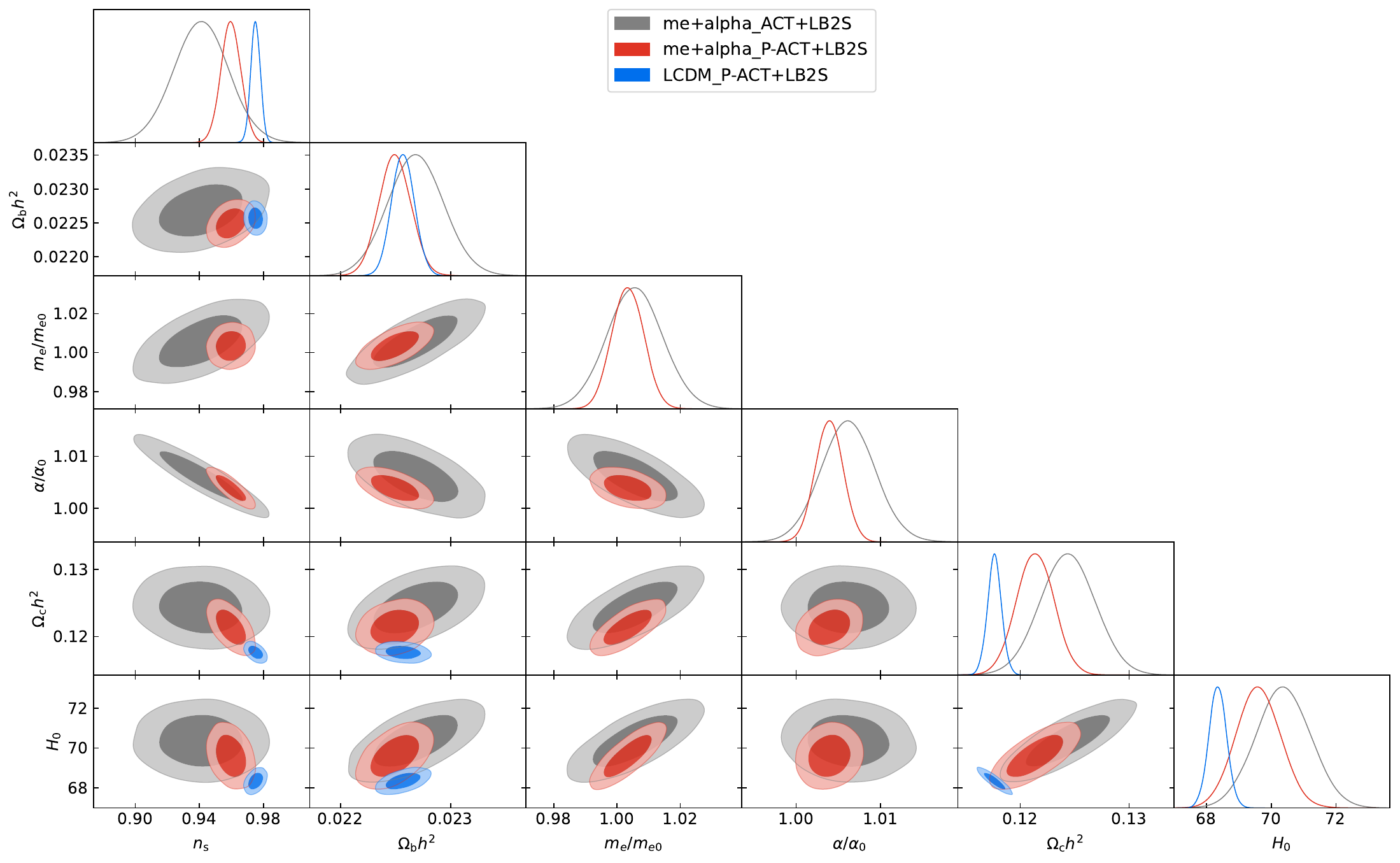} 
\centering
\caption{Posterior distributions of selected parameters for the varying $m_e$\&$\alpha$ model. Gray and red contours correspond to the different datasets indicated in the legend, while blue contours represent the $\Lambda$CDM baseline for comparison.}
\label{fig:me-alpha} 
\end{figure}

In Table~\ref{table} and Figure~\ref{fig:me-alpha}, we also show the constraints on the simultaneous variation of $\alpha$ and $m_e$.
Such a scenario could be realized in certain
class of models~\cite{Brax:2002nt,Chiba:2006xx,Barrow:2011kr}. 
In this case, instead of the electron mass approaching its standard value, the fine-structure constant deviates from the standard value, which differs from the result obtained using Planck+LB2S (see Ref.~\cite{Toda:2025dzd} for the Planck+LB2S result and the implementation of the varying $\alpha$ model).
If we use 6df and SDSS instead of DESI BAO, we obtain
\begin{align}
\begin{cases}
m_e/m_{e0}=0.9982\pm 0.0057\\
\alpha/\alpha_{0}=1.0038\pm 0.0017\\
H_0=68.34\pm 0.97\\
n_\mathrm{s}=0.9594\pm 0.0062\\
\end{cases} \quad (\mathrm{P\!-\!ACT\!-\!LB_{6S}S}). 
\end{align}
which indicates $m_e/m_{e0}=1$ is well consistent, while a little larger fine structure constant is preferred.
It is also worth noting that simultaneous variations in $\alpha$ and $m_e$ further reduce Hubble tension than the variation of only $m_e$, both in terms of AIC and Gaussian tension.

\subsection{Early Dark Energy}

\begin{figure}[ht]
\includegraphics[width=17cm]{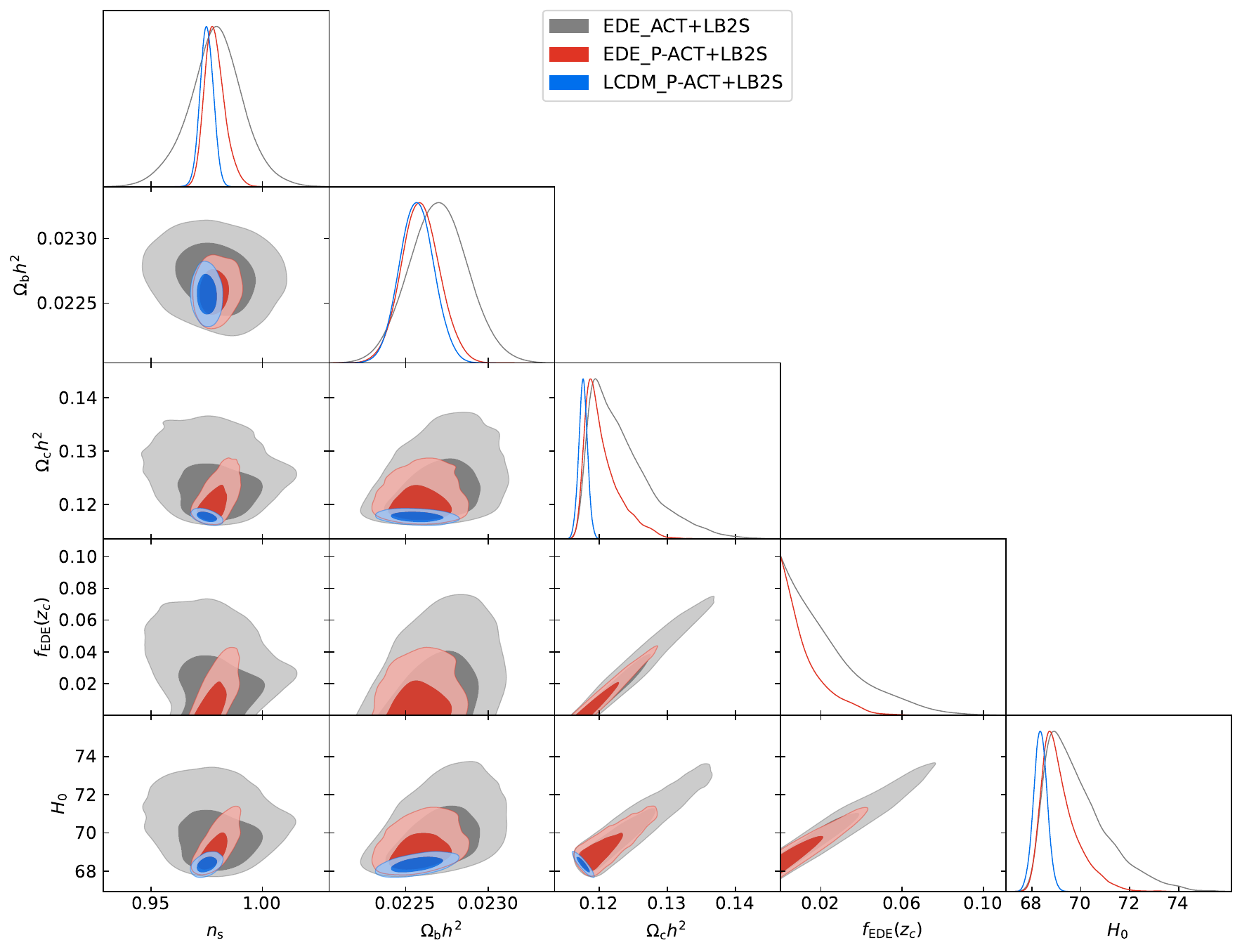} \caption{Posterior distributions of selected parameters for the EDE model. Gray and red contours correspond to the different datasets indicated in the legend, while blue contours represent the $\Lambda$CDM baseline for comparison.}
\label{fig:EDE} 
\end{figure}

\begin{figure}[ht]
\includegraphics[width=16.7cm]{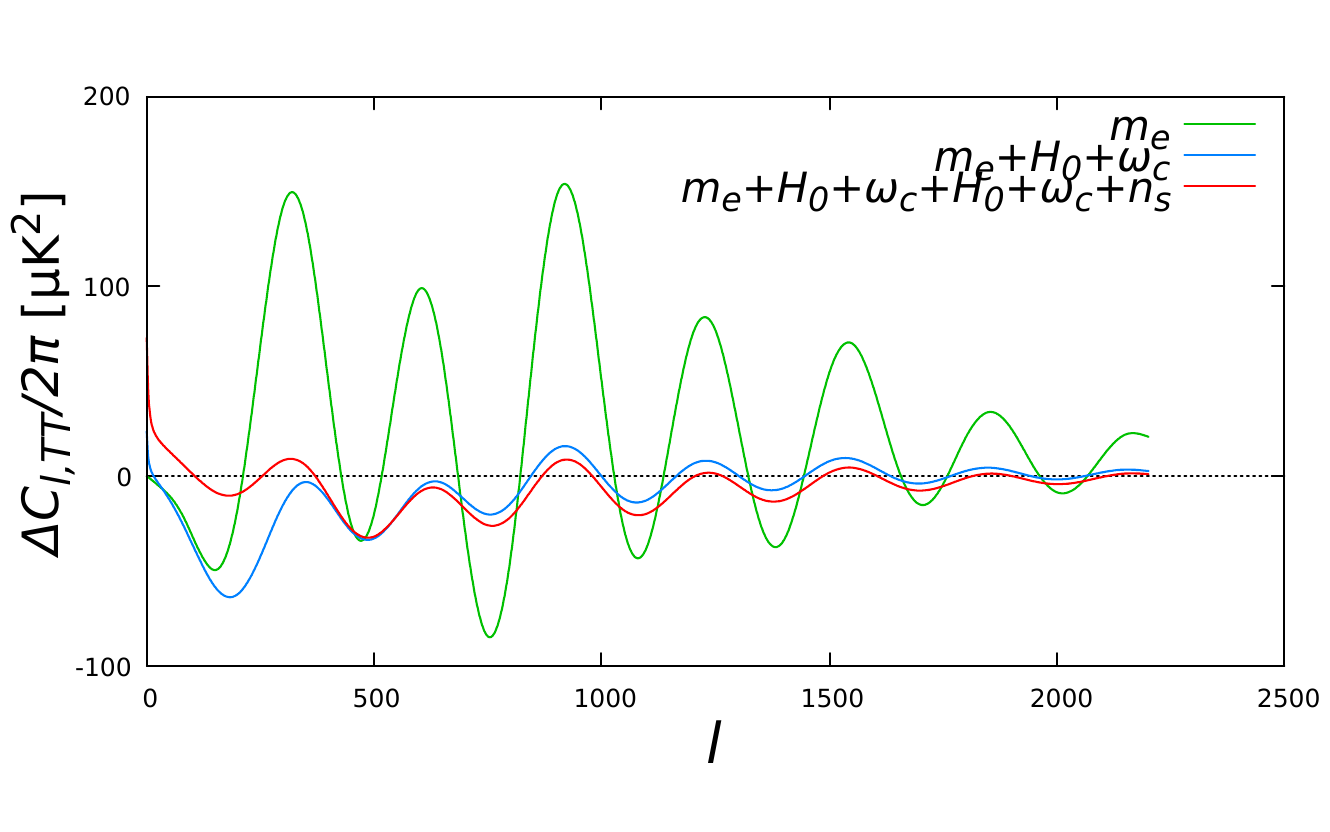}
\includegraphics[width=16.7cm]{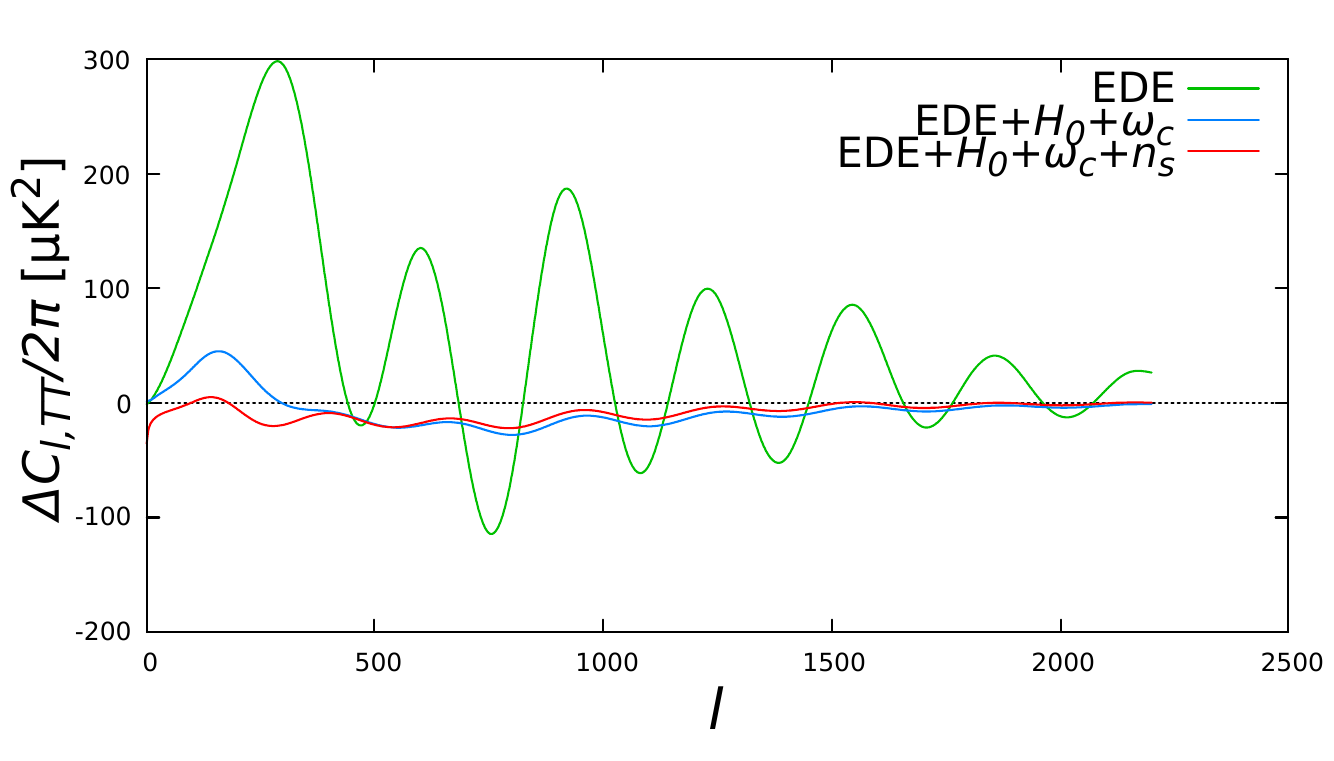}
\centering
\caption{The CMB power spectrum difference relative to the $\Lambda$CDM model. 
The upper green curve shows the case with only a $2\%$ increase in the electron mass. 
The blue curve additionally includes variations in $H_0$, $\omega_c$, and $\omega_b$ and the red curve further includes a variation in $n_s$ according to Eq.~(\ref{eq.mefit}).
The bottom panel shows the same comparison but for the EDE model according to Eq.~(\ref{eq.EDEfit}).}
\label{fig:TT} 
\end{figure}

\begin{figure}[ht]
\includegraphics[width=16.7cm]{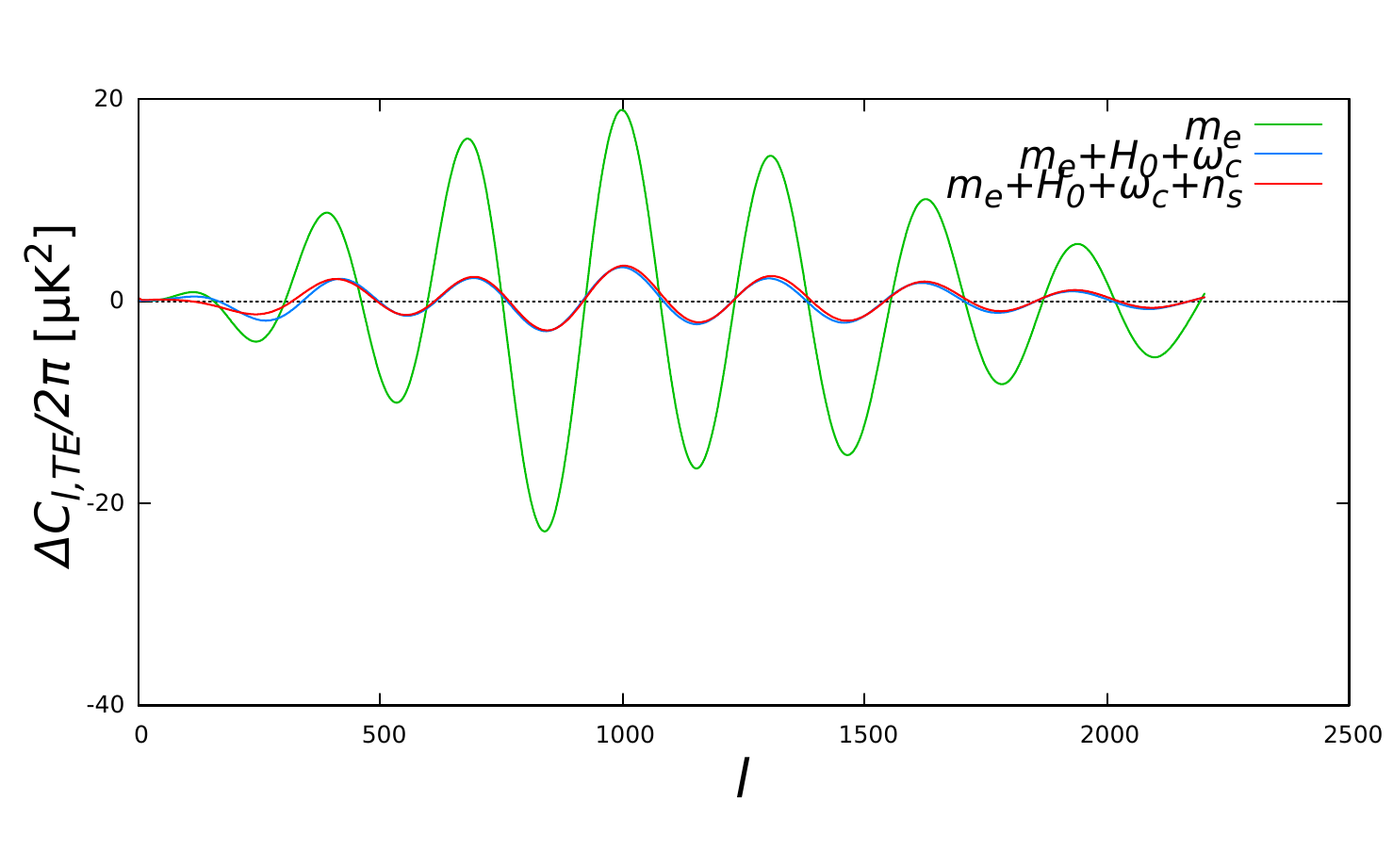}
\includegraphics[width=16.7cm]{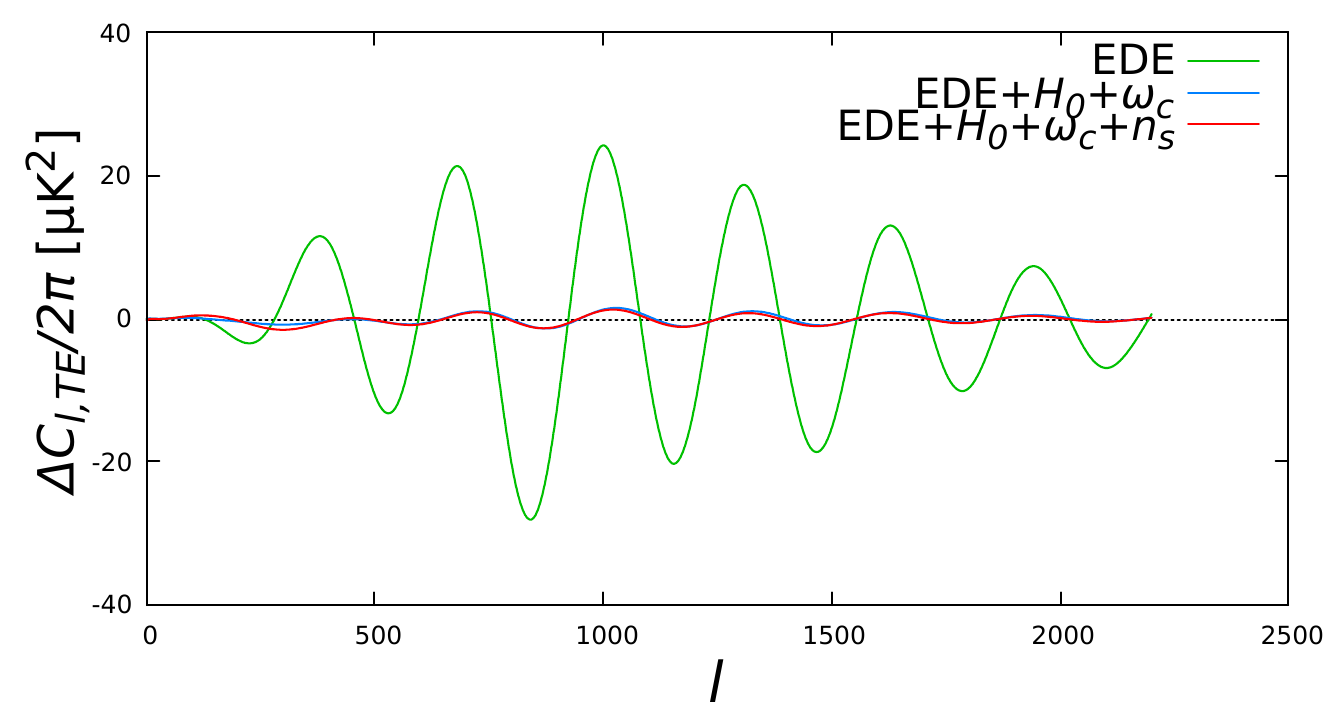}
\centering
\caption{Same figure as Fig.~\ref{fig:TT} for $C_{l, TE}$}
\label{fig:TE} 
\end{figure}

We find that a larger amount of EDE, characterized by $f_{\mathrm{EDE}}$, leads to a higher Hubble constant and relieves the Hubble tension.
However, for the ACT+LB2S and P-ACT+LB2S analyses, we do not find a lower bound on $f_{\mathrm{EDE}}$ 
or preference of the existence of EDE.
This result is consistent with the previous work~\cite{Poulin:2025nfb} which examined the case of $n=3$.
Similar to the varying $m_e$ model, Fig.~\ref{fig:EDE} shows the approximate relation for the fractional changes\footnote{See also Ref.~\cite{Jiang:2022uyg} for other EDE models. The coefficients do not differ significantly.}:
\begin{equation}
    f_{\mathrm{EDE}} \simeq 
    \frac{1}{2} \frac{\Delta \omega_\mathrm{c}}{\omega_\mathrm{c}} 
    \simeq \frac{8}{7} \frac{\Delta H_0}{H_0}
    \simeq 5 \frac{\Delta n_s}{n_s}.
    \label{eq.EDEfit}
\end{equation}
It is shown in the bottom panel of Fig.~\ref{fig:TT},\ref{fig:TE} that how this degeneracy recoveres the CMB power spectrum relative to the $\Lambda$CDM model.

From the perspective of the Hubble tension, we find that the EDE model also provides an effective solution, reducing the tension to $3.62\sigma$, which is $1.73\sigma$ lower than that for the standard $\Lambda$CDM model.
However, this is less effective than the varying $m_e$ model.
The best-fit total chi-squared value, $\chi^2_{\mathrm{total}}$, is improved by 7.38, which is slightly larger than that of the varying $m_e$ model.
However, since the EDE model introduces three additional free parameters compared to $\Lambda$CDM, the resulting $\Delta \mathrm{AIC} = -1.38$ indicates that this model alleviates the Hubble tension less effectively.

\subsection{Implication for the inflation}

The predictions of inflation models are primarily expressed by two major parameters: the scalar spectral index $n_s$ and the tensor-to-scalar ratio $r$.
As shown in Figs.~\ref{fig:me} and~\ref{fig:EDE} and Eqs.~(\ref{eq.mefit}) and~(\ref{eq.EDEfit}), 
a larger electron mass $m_e / m_{e0}$ prefers a smaller $n_s$, while a larger EDE fraction $f_{\mathrm{EDE}}$ does a larger $n_s$ even though both scenarios lead to a higher $H_0$.
Therefore, depending on which scenario is adopted to address the Hubble tension,
 a different inflation model seems to be suitable.

\begin{figure}[ht]
\includegraphics[width=17cm]{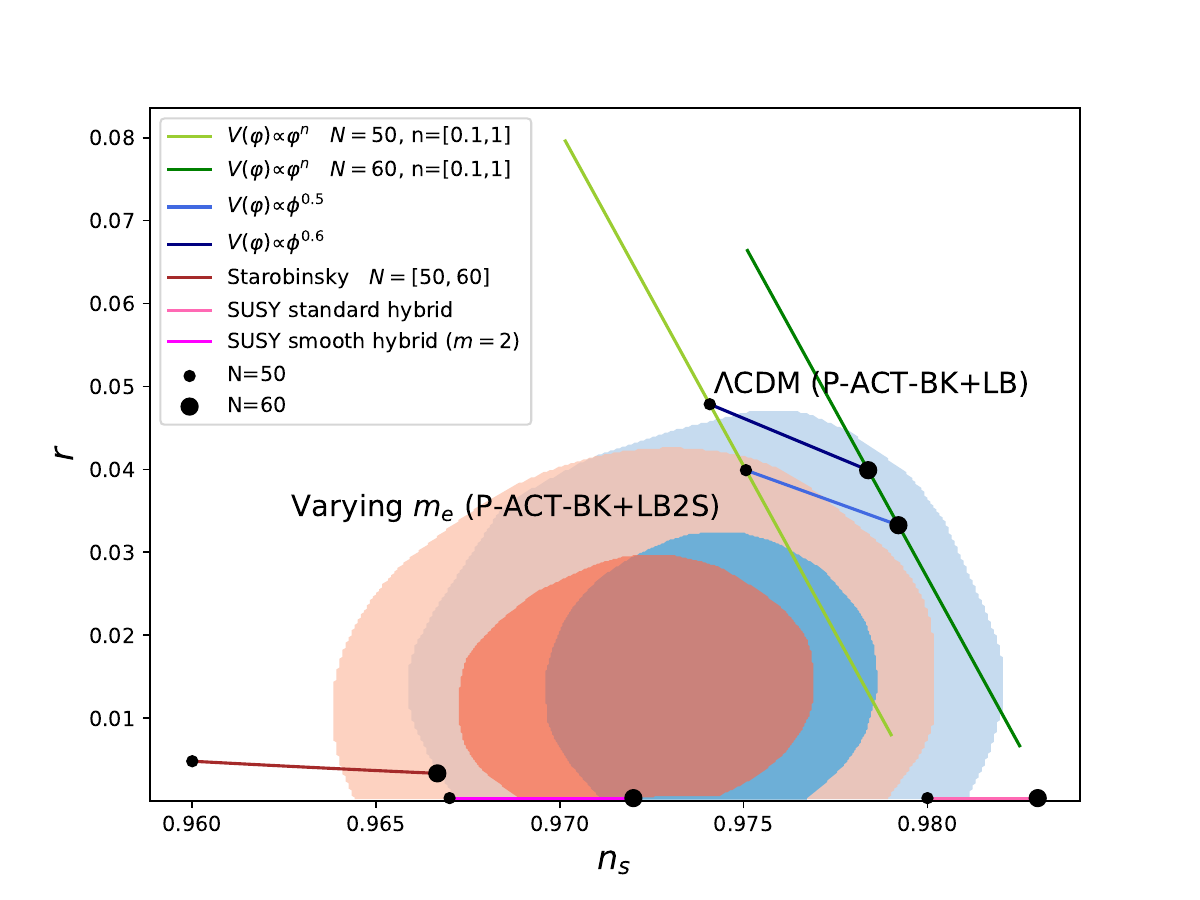} 
\centering
\caption{Constraints on $n_s$ and $r$. 
The orange contour represents the varying $m_e$ case using the P-ACT-BK+LB2S dataset, while the blue contours show the $\Lambda$CDM baseline for comparison from Ref.~\cite{ACT:2025tim}. 
The green and yellow-green lines indicate predictions for different power-law potentials with the number of e-folds of inflation $N=60$ and $N=50$. 
The purple line indicates prediction for the Starobinsky model with the number of e-folds $N=60$ to $N=50$. 
}
\label{fig:Inflation} 
\end{figure}

\begin{figure}[ht]
\includegraphics[width=17cm]{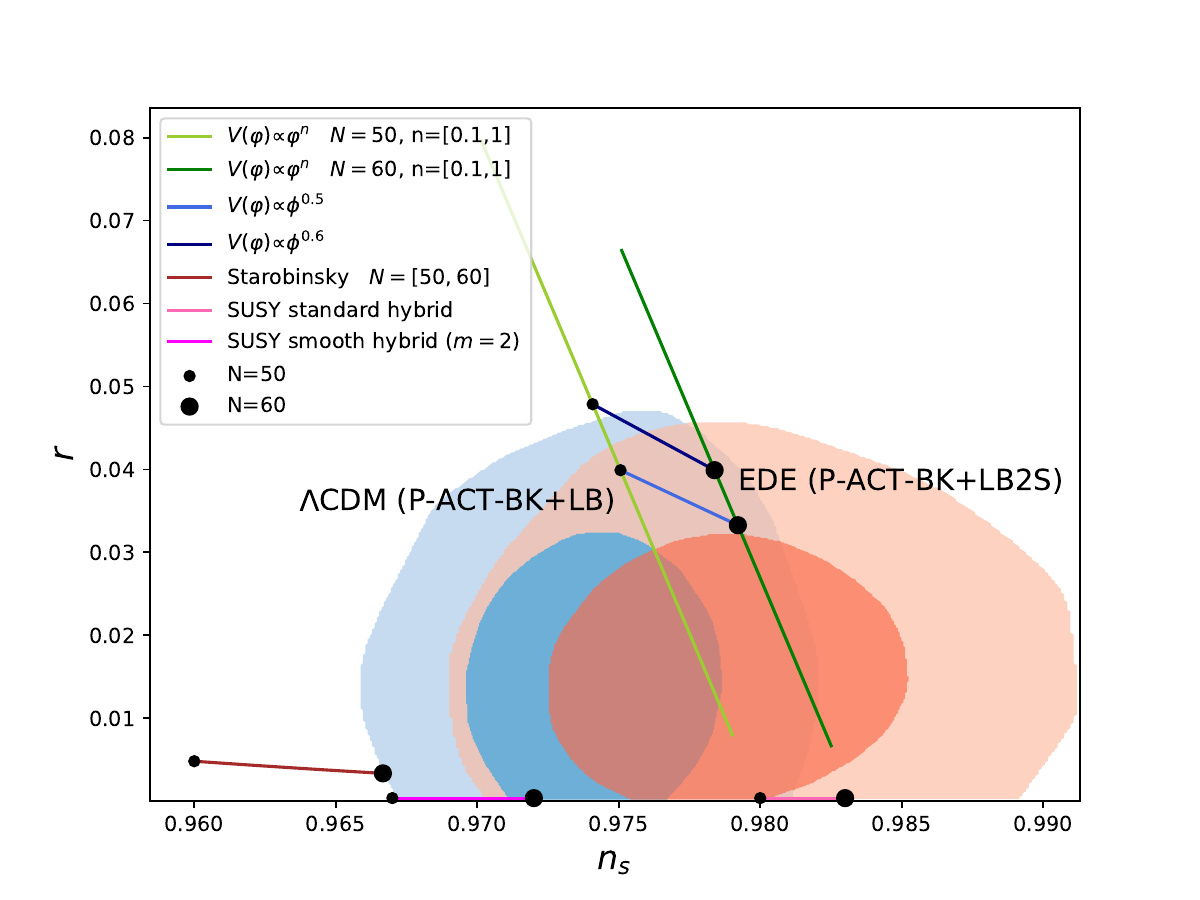} 
\centering
\caption{Same figure as Fig.~\ref{fig:Inflation} for EDE.
}
\label{fig:Inflation2} 
\end{figure}

For the varying $m_e$ model, we find $n_s = 0.9717 \pm 0.0068$ (ACT-LB2S) and $n_s = 0.9721 \pm 0.0035$ (P-ACT-LB2S), both of which are lower than the values in the standard $\Lambda$CDM model.
A two-dimensional constraint on $n_s$ and $r$ for the varying $m_e$ model, incorporating the BICEP-Keck Array data, is shown in Fig.~\ref{fig:Inflation}.
This lower value of $n_s$ brings the Starobinsky inflation model with $N=60$~\cite{Starobinsky:1980te}
 as well as the potential $V(\phi) \propto \phi^{n}$ with $N=50$ and $n \lesssim 0.5$ lies within the $2\sigma$ region.
One of the best fit model would be ``smooth hybrid inflation''~\cite{Lazarides:1995vr} that is a variation of supersymmetric hybrid inflation with the superpotential
 $ W =S \left( \mu-\frac{(\bar{\Psi}\Psi)^m}{M^{m-2}} \right) $, where $S$ and $\Psi$ are superfields contain the inflaton and the second field respectively, and $\mu$ being constant.
We also display the prediction of smooth hybrid inflation with the power of nonrenormalizable term $m=2$ in Figs.~\ref{fig:Inflation} and \ref{fig:Inflation2}. Other case with a larger $m$ can be found in Ref.\cite{Okada:2025lpl}. 

For the EDE scenario, we find $n_s = 0.979 \pm 0.013$ (ACT-LB2S) and $n_s = 0.9787^{+0.0036}_{-0.0051}$ (P-ACT-LB2S), both of which are higher than the values in the standard $\Lambda$CDM model.
The former even reaches $n_s = 1$ within the $2\sigma$ level.
A two-dimensional constraint on $n_s$ and $r$ for EDE, using BICEP-Keck Array data, is shown in Fig.~\ref{fig:Inflation2}.
The smooth hybrid inflation with $N=60$~\cite{Starobinsky:1980te} as well as
the potential $V(\phi) \propto \phi^{n}$ with $n \lesssim 0.6$ lies within the $2\sigma$ region.
This higher value of $n_s$ brings the supersymmetric inflation model~\cite{Dvali:1994ms}, which predicts $n_s \simeq 0.98$ for $N=50$, within the $1\sigma$ region. Although this original model nowadays suffers from the formation of topological defects, modifications to avoid this problem without altering the prediction is also possible~\cite{Jeannerot:2000sv}.

\begin{table*}[h]
\[
\begin{tabular}{lcccc}
\hline\hline 
Parameter  &  \ensuremath{\Lambda}CDM  &  varying\,\ensuremath{m_{e}}  & varying\,\ensuremath{m_{e}\,\&\,\alpha}  &  EDE\\
\hline  {\boldmath\ensuremath{\Omega_{b}h^{2}}}  & \ensuremath{0.02262\pm0.00016} & \ensuremath{0.02298\pm0.00021} & \ensuremath{0.02268\pm 0.00025} & \ensuremath{0.02270\pm0.00018}\\
 {\boldmath\ensuremath{\Omega_{c}h^{2}}}  & \ensuremath{0.11785\pm0.00072} & \ensuremath{0.1244\pm0.0026} & \ensuremath{0.1244\pm 0.0025} & \ensuremath{0.1234_{-0.0056}^{+0.0019}}\\
 {\boldmath\ensuremath{m_{e}/m_{e0}}}  &  & \ensuremath{1.0174\pm0.0065} & \ensuremath{1.0056\pm 0.0088} \\
 {\boldmath$\alpha/\alpha_{0}$} &  &  & \ensuremath{1.0062\pm 0.0032}\\
 {\boldmath\ensuremath{f_{\mathrm{EDE}}(z_{c})}}  &  & &  & \ensuremath{<0.0268}\\
 {\boldmath\ensuremath{n_{s}}}  & \ensuremath{0.9764\pm0.0067} & \ensuremath{0.9717\pm0.0068} & \ensuremath{0.941\pm 0.017} & \ensuremath{0.979\pm0.013}\\
\ensuremath{\Omega_\mathrm{m}} & \ensuremath{0.3020\pm 0.0039} & \ensuremath{0.2978\pm 0.0041} & \ensuremath{0.2983\pm 0.0041 } & \ensuremath{0.3004\pm 0.0041}\\

 \ensuremath{H_{0}}  & \ensuremath{68.36\pm0.29} & \ensuremath{70.50\pm0.86} & \ensuremath{70.37\pm 0.85} & \ensuremath{69.89_{-1.6}^{+0.60}}\\
 \ensuremath{\sigma_{8}}  & \ensuremath{0.8130_{-0.0055}^{+0.0049}} & \ensuremath{0.841\pm0.012} & \ensuremath{0.829\pm 0.013} & \ensuremath{0.8194_{-0.0088}^{+0.0071}}\\
\hline  \hline
\end{tabular}
\]

\[
\begin{tabular}{lcccc}
 \hline\hline
 Parameter  &  \ensuremath{\Lambda}CDM  &  varying\,\ensuremath{m_{e}} & varying\,\ensuremath{m_{e}\,\&\,\alpha} &  EDE\\
\hline  {\boldmath\ensuremath{\Omega_{b}h^{2}}}  &  \ensuremath{0.02257\pm0.00010} & \ensuremath{0.02269\pm0.00012} & \ensuremath{0.02249\pm 0.00014} & \ensuremath{0.02259\pm0.00012}\\
 {\boldmath\ensuremath{\Omega_{c}h^{2}}}  &  \ensuremath{0.11765\pm0.00065} & \ensuremath{0.1202\pm0.0016} & \ensuremath{0.1214\pm 0.0017} & \ensuremath{0.1206_{-0.0032}^{+0.0010}}\\
 {\boldmath\ensuremath{m_{e}/m_{e0}}}  &   & \ensuremath{1.0078\pm0.0047}& \ensuremath{1.0034\pm 0.0050}\\
 {\boldmath$\alpha/\alpha_{0}$} & & & \ensuremath{1.0039\pm 0.0016}          \\
 {\boldmath\ensuremath{f_{\mathrm{EDE}}(z_{c})}}  &   &  &  & \ensuremath{<0.0140}\\
 {\boldmath\ensuremath{n_{s}}}  &  \ensuremath{0.9750\pm0.0030} & \ensuremath{0.9721\pm0.0035} & \ensuremath{0.9596\pm 0.0062} & \ensuremath{0.9787_{-0.0051}^{+0.0036}}\\
 {\ensuremath{\Omega_\mathrm{m}}}   & \ensuremath{0.3015\pm 0.0036}   & \ensuremath{0.2979\pm 0.0040 } &   \ensuremath{0.2984\pm 0.0040 }    & \ensuremath{0.3006\pm 0.0036}          \\
 \ensuremath{H_{0}}  &  \ensuremath{68.35\pm0.27} & \ensuremath{69.41\pm0.68} & \ensuremath{69.60\pm 0.69} & \ensuremath{69.17_{-0.90}^{+0.38}}\\
 \ensuremath{\sigma_{8}}  &  \ensuremath{0.8123_{-0.0051}^{+0.0045}} & \ensuremath{0.8239\pm0.0082} & \ensuremath{0.8227\pm 0.0083} & \ensuremath{0.8165_{-0.0068}^{+0.0054}}\\
\hline 
 \ensuremath{T_{H_{0}}} & 5.35\ensuremath{\sigma} & 3.43\ensuremath{\sigma} & 3.24\ensuremath{\sigma} & 3.62\ensuremath{\sigma}\\
 \ensuremath{T_{\sigma_{8}}}  & 0.50\ensuremath{\sigma} & 1.01\ensuremath{\sigma} & 0.95\ensuremath{\sigma} & 0.69\ensuremath{\sigma} \\
 \hline\hline
\end{tabular}
\]

\caption{68\% constraints from ACT+LB2S (top) and P-ACT+LB2S (bottom) and the gaussian tension to other measurements
\label{table}}
\end{table*}

\begin{table*}[h]
\[
\begin{tabular}{lcccc}
 \hline\hline
 Parameter  &  \ensuremath{\Lambda}CDM  &  varying\,\ensuremath{m_{e}} & varying\,\ensuremath{m_{e}\,\&\,\alpha} &  EDE\\
\hline  {\boldmath\ensuremath{\Omega_{b}h^{2}}}  & 0.0226076 & 0.0227507 & 0.0225862 & 0.0226207\\
 {\boldmath\ensuremath{\Omega_{c}h^{2}}}   & 0.117136 & 0.120794 & 0.123188 & 0.125382\\
 {\boldmath\ensuremath{m_{e}/m_{e0}}}  & & 1.01053 & 1.00857 & \\
 {\boldmath$\alpha/\alpha_{0}$} & & & 1.0047 & \\
 {\boldmath\ensuremath{f_{\mathrm{EDE}}(z_{c})}}  & & & & 0.0327615\\
 {\boldmath\ensuremath{n_{s}}} & 0.976315 & 0.972152 & 0.955113 & 0.983925\\
 \ensuremath{H_{0}}  & 68.5579 & 69.8702 & 70.6637 & 70.6517\\
 \ensuremath{\sigma_{8}}  & 0.812026 & 0.827638 & 0.829986 & 0.820289\\
\hline 
 \ensuremath{\chi^2_{\mathrm{BAO}}}  & 11.04 & 10.33 & 10.91 & 11.18\\
 \ensuremath{\chi^2_{\mathrm{Planck}}}  & 635.22 & 635.18 & 634.45 & 633.78\\
 \ensuremath{\chi^2_{\mathrm{ACT}}} & 174.30 & 171.33 & 167.42 & 173.53\\
 \ensuremath{\chi^2_{\mathrm{Mb}}}  & 7.13 & 3.43 & 2.04 & 1.92\\
 \ensuremath{\chi^2_{\mathrm{SN}}}  & 1034.86 & 1034.98 & 1034.84 & 1034.75\\
 \hline
 \ensuremath{\chi^2_{\mathrm{total}}} & 1862.55 & 1855.23 & 1849.65 & 1855.17\\
 \ensuremath{\chi^2_{\mathrm{total}}-\chi^2_{\mathrm{total, \, \Lambda CDM}}} & 0.00 & -7.32 & -12.90 & -7.38\\
 \ensuremath{\Delta AIC} & 0.00 & -5.32 & -8.90 & -1.38\\
 \hline\hline
\end{tabular}
\]
\caption{The best-fit values of the cosmological parameters and their corresponding minimized chi-squared values. We use the data P-ACT+LB2SMb. 
\label{table-best}}
\end{table*}

\section{Conclusions}
\label{sec:conclusions} 

In this paper, we have examined the varying electron mass model and the axionlike EDE model, taking account of the recent ACT CMB measurements.
The previous analyses using the latest Planck CMB and DESI BAO data have shown that a slightly larger electron mass, $m_{e} / m_{e0} = 1.0101 \pm 0.0046$ (Planck+LB2S), is preferred and the standard value $m_{e}/m_{e0} = 1$ is more than $2\sigma$ away~\cite{Toda:2025dzd}.
This analysis including ACT data indicates that the mean value of $m_e/m_{e0}$ is about $\sim 1\%$ larger value of the electron mass as
\begin{align}
&\begin{cases}
m_{e} / m_{e0} = 1.0078 \pm 0.0047 \\
H_{0} = 69.41 \pm 0.68~\mathrm{km\,s^{-1}Mpc^{-1}} \\
n_s = 0.9721 \pm 0.0035
\end{cases} \quad (\mathrm{P\!-\!ACT\!-\!LB2S}),\\
&\begin{cases}
m_{e} / m_{e0} = 1.0081 \pm 0.0046 \\
H_{0} = 69.42 \pm 0.68~\mathrm{km\,s^{-1}Mpc^{-1}} \\
n_s = 0.9721 \pm 0.0035 \\
r = 0.0164^{+0.0066}_{-0.013}
\end{cases} \quad (\mathrm{P\!-\!ACT\!-\!BK\!-\!LB2S}).
\end{align}
It is worth emphasizing that the ACT data, which provide high-precision measurements of the CMB power spectrum in the high-$l$ multipole range, support this slight preference for a larger electron mass.
We also clarify that the varying electron mass model leads to a lower spectral index $n_s$, which brings the Starobinsky inflation model within the $2\sigma$ region and smooth hybrid inflation well fits with data.

In parallel, we have also investigated the axionlike EDE model to assess how it compares in addressing the Hubble tension and its impact on inflationary parameters.
In contrast to the varying electron mass model, we do not find significant favor of the existence of EDE:
\begin{align}
&
\begin{cases}
f_{\mathrm{EDE}}(z_c) < 0.0140 \\
H_{0} = 69.17^{+0.38}_{-0.90}~\mathrm{km\,s^{-1}Mpc^{-1}} \\
n_s = 0.9787^{+0.0036}_{-0.0051}
\end{cases} 
\quad (\mathrm{P\!-\!ACT\!-\!LB2S}),\\
&
\begin{cases}
f_{\mathrm{EDE}}(z_c)< 0.0144 \\
H_0 = 69.17^{+0.38}_{-0.92}  \\ 
n_s=0.9790^{+0.0037}_{-0.0051} \\
r=0.0176^{+0.0072}_{-0.013} \\
\end{cases} 
\quad (\mathrm{P\!-\!ACT\!-\!BK\!-\!LB2S}).
\end{align}
The EDE scenario leads to a higher spectral index $n_s$, which brings the supersymmetric hybrid inflation model within the $2\sigma$ region rather than the Starobinsky model.
Therefore, the choice of solution for the Hubble tension directly affects the choice of inflationary models.

\begin{acknowledgments}
\noindent This work O.S. was in part supported by JSPS KAKENHI Grant No. 23K03402. 
\end{acknowledgments}

\bibliography{me_DESI}

\end{document}